\def\nh{{n_{\rm H}}}
\def\nh2{{n(\rm H_2)}}
\def\h2{${\rm H_2}$}

\def\3cm{\rm {cm^{-3}}}
\def\2cm{\rm {cm^{-2}}}
\def\s-1{\rm {s^{-1}}}

\def\mum {\hbox{$\mu$m}}
\def\kms {\hbox{${\rm km\,s}^{-1}$}}
\def\Kkms {\hbox{${\rm K}\,{\rm km}\,{\rm s}^{-1}$}}

\def\twco{\hbox{$^{12}$CO}}
\def\thco{\hbox{$^{13}$CO}}

\def\ci{\hbox{\rm {[C {\scriptsize I}]}}}
\def\cii{\hbox{\rm {[C {\scriptsize II}]}}}

\def\c18o{\hbox{C$^{18}$O}}

\documentclass[letter]{aa} 
\usepackage{graphicx}
\usepackage{epsfig}
\usepackage[]{natbib}
\usepackage{txfonts}
\usepackage{psfrag}
\usepackage{units}
\begin{document}
\title{[CII] gas in IC~342}
\author{M. R\"ollig\inst{1} \and 
        R.~Simon\inst{1} \and
        R.~G\"usten\inst{2} \and
        J.~Stutzki\inst{1} \and
        H.~W. H\"ubers\inst{3,4} \and
        P.~Hartogh\inst{5} \and
	K.~Jacobs\inst{1} \and
	X.~Guan\inst{1}\and
        F.~Israel\inst{6}
}
\offprints{M. R\"ollig}
\institute{
 I. Physikalisches Institut der Universit\"at zu K\"oln, Z\"ulpicher Stra\ss e 77, 50937 K\"oln, Germany\\
 \email{roellig@ph1.uni-koeln.de} 
\and
 Max-Planck-Institut f\"ur Radioastronomie, Auf dem H\"ugel 69, 53121 Bonn, Germany 
\and
Deutsches Zentrum f\"ur Luft- und Raumfahrt, Institut f\"ur Planetenforschung,   
          Rutherfordstra\ss e 2, 12489 Berlin / Germany
\and
Institut f\"ur Optik und Atomare Physik, Technische Universit\"at Berlin, 
         Hardenbergstra\ss e 36, 10623 Berlin / Germany
\and
Max-Planck-Institut f\"ur Sonnensystemforschung, 
      Max-Planck-Stra\ss e 2, 37191 Katlenburg-Lindau/ Germany
\and
Leiden Observatory, Leiden University, PO Box 9513, NL 2300 RA Leiden, The Netherlands
}
\date{Received  / Accepted  }
%

%
\abstract
  {
  }
  { 
  }
  {We used the dual-band receiver GREAT on board the SOFIA airborne telescope to perform observations of the \cii~158~\mum\ fine-structure line at the  postitions of two giant molecular clouds (GMC) in the center of IC~342 (GMCs C and E)
  and compared the spectra  with corresponding ground-based data for low- and mid-$J$ CO and \ci{}. We performed model calculations assuming a clumpy photo-dissociation region (PDR) environment using the KOSMA-$\tau$ PDR model code to derive physical parameters of the local medium. 
  }
  {The \cii~158~\mum\ emission resembles the spectral signature of ground-based atomic and molecular lines, which indicates a common origin. The emission from GMC E can be decomposed into a cool, molecular component with weak far-ultraviolet (FUV) fields and low, mean densities of 10$^3$~cm$^{-3}$ and a strongly excited starburst/PDR region with higher densities of 10$^4$~cm$^{-3}$ and FUV intensities of 250-300 Draine fields. The emission from GMC C  is consistent with gas densities of 5000~cm$^{-3}$, FUV intensities of a few Draine fields and total gas masses of 20 $\times 10^6$~M$_\odot$.
  }
  {
The high spectral resolution of the GREAT receiver allowed us to decompose the \cii\ emission of the GMC E into a strongly excited gas component resembling a PDR/starburst environment and a quieter, less excited gas component and to analyze the different components within a single beam individually.   
  }
  
\keywords{galactic: ISM
--- galactic: individual: IC 342
--- radio lines: extragalactic
--- radio lines: ISM
--- atoms: [C II]}

\maketitle

\section{Introduction}
IC~342 is a gas-rich spiral galaxy with active star formation in its nucleus. IC~342 is located behind the Galactic plane und is therefore highly obscured; the distance to the nearly face-on galaxy is still debated. \citet{tikhonov2010} give a distance of $3.9\pm0.1$~Mpc derived from stellar photometry. Observations of planetary nebulae \citep{herrmann2008} and cepheids \citep{saha2002} give distances of 3.4$\pm$0.2 Mpc.
   
\begin{figure}[!tp]

\epsfig{file=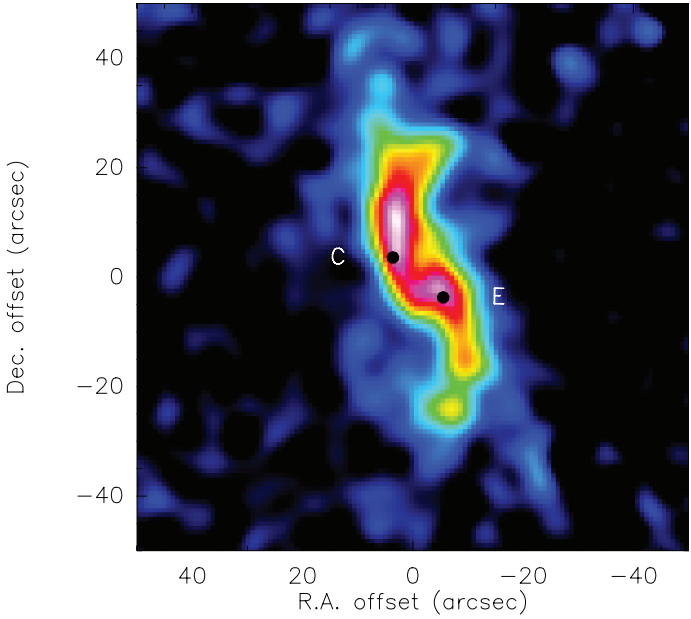,angle=0,width=1\linewidth}

  \caption{\footnotesize{Line-integrated map of the \twco(1-0) transition from the BIMA-SONG sample. The black dots denote the postitions of our {\bf two} \cii\ detections in IC~342. The positions correspond to giant molecular clouds C and E. The (0,0) position corresponds to (RA,DEC) (J2000) (03:46:48.5 68:05:47).}}
  \label{fig:1}
\end{figure}

Within its central 30$''$ two molecular arms of a mini-spiral end in a clumpy central ring of dense gas, which surrounds a young star cluster. \citet{downes1992} showed the presence of five giant molecular clouds, A to E, around the nucleus of IC~342 with masses of $\sim 10^6$~M$_\odot$. This structure can be seen in Fig. \ref{fig:1}, which shows a map of line-integrated \twco(1-0) emission observed with BIMA
(Berkeley Illinois Maryland Association).
The central molecular ring surrounds a nuclear star cluster with active star formation and strong far-ultraviolett (FUV) radiation, illuminating the molecular ring and producing photo-dissociation regions (PDRs) on the side facing the central cluster. Energetiv FUV photons dissociate and ionize molecules and atoms in the gas and effectively heat the gas and dust via photoelectric heating. Consequently, PDRs strongly emit radiation from species that are abundant and excited under these conditions, like \cii. 
\citet{meier2005} find that the chemistry in the ring is a mixture of PDR 
gas and regions of denser, more shielded material.
Especially the GMC complex B appears to show properties of PDRs produced by the surrounding starburst environment \citep{downes1992,israel2003}. 
The spatial sizes of the central GMCs and the infrared luminosity of the inner 400 pc of IC~342 are both similar to the center of our Galaxy. 
Especially the relative contribution of the diffuse material to the overall \cii\ emission remains unknown but it appears that the PDRs  remain largely confined to the central ring.

Detailed knowledge of the H$_2$ column densities, excitation temperatures, and densities in IC 342 exists from CO and its isotopomers as a basis for comparison with the \cii\ lines \citep[e.g.][]{ishizuki1990,downes1992,turner1992,wright1993,turner1993,meier2000,meier2001,israel2003,meier2005}. 
Comparison with this complementary data, particularly with their spectral shape, allows us to distinguish different kinematic components from the \cii\ lines and to assess the interaction of the star formation activity on the gas.

In this work we present 
velocity-resolved 
spectra of ionized atomic (\cii~158~\mum) gas, with spatial resolutions of $\sim$15$''$.6.
The observations were performed with the German Receiver for Astronomy at Terahertz Frequencies\footnote{GREAT is a development by the MPI f\"ur Radioastronomie and 
KOSMA/ Universit\"at zu K\"oln, in cooperation with the MPI f\"ur Sonnensystemforschung and the DLR Institut f\"ur Planetenforschung.
} \citep[GREAT,][]{GREAT} 
on board the Stratospheric Observatory For Infrared Astronomy (SOFIA).

In Sect. 2 we describe the observations. The spectra are presented in Sect. 3. The analysis of the data and ambient conditions are discussed in Sect. 4, and conclusions are drawn in Sect. 5.




\section{Observations}

We used the dual-channel receiver GREAT on SOFIA in September 2011
to perform pointed observations of the \cii~158~\mum\ fine-structure transition of C$^+$  at 1900.536900 GHz and the \twco(11-10) transition at 1496.922909 GHz 
near 
the center of IC~342.
The observations were performed in dual beam-switch mode (chop rate 1~Hz) toward selected positions, see Tab.~\ref{tab:1} for details. The chopper offset was 100$''$ with a chopper angle of 
110 degrees against north (counter-clockwise).
 The observations were taken during the transfer flight from the US to Europe. Owing to unknown technical errors on the flight we had to discard one of the observed positions.

The center position is RA,DEC (J2000) 03:46:48.5 68:05:47. We observed 2 GMCs at the following offsets in arcsecond: GMC C (+3.6,+3.6), and GMC E (-5.5,-3.7) using the designation by \citet{downes1992} (see also Fig.~\ref{fig:1}). The integration times were 9.3, and 3.7 minutes, respectively.
 \cii\ emission was detected at both positions. We did not detect any \twco(11-10) emission. The rms of the baseline for position C and E was 0.10 and 0.21 \Kkms\ , respectively, with a channel width of 4 \kms. 

We used a fast Fourier transform spectrometer (FFTS), with 8192 channels providing 1.5 GHz bandwidth and about 212 kHz of spectral resolution. 
Calibration was performed with the standard pipeline \citep{guan2012}.
Using the beam efficiency $\eta_c\approx0.51$ and the forward efficiency ($\eta_f$) of 0.95 \citep{GREAT},
we converted all data to line brightness temperature scale, $T_{B}=\eta_{f}\times T_{A}^{*}/\eta_{c}$. 
The reduction of these calibrated data, as well as the maps shown throughout the paper, were made with the GILDAS\footnote{http://www.iram.fr/IRAMFR/GILDAS} package CLASS90. The pointing was established with the optical guide cameras to about 1$''$ precision, and was found to have deviated
 by about 3-5$''$ after 20 minutes.
\section{Results}

 \begin{figure}[!tp]
 \centering  
 \begin{tabular}{c}
   \epsfig{file=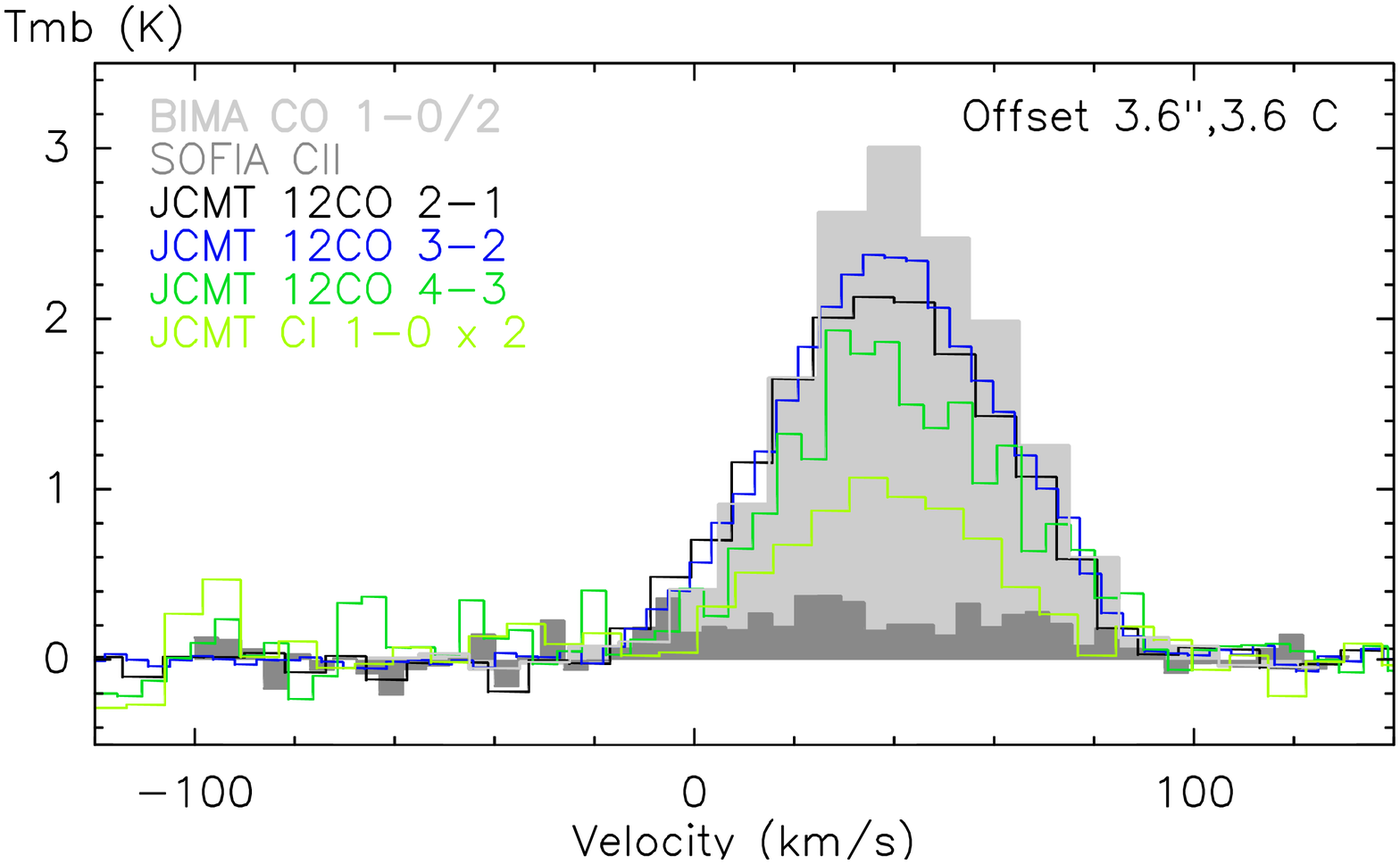,angle=0,width=0.85\linewidth} \\
   \epsfig{file=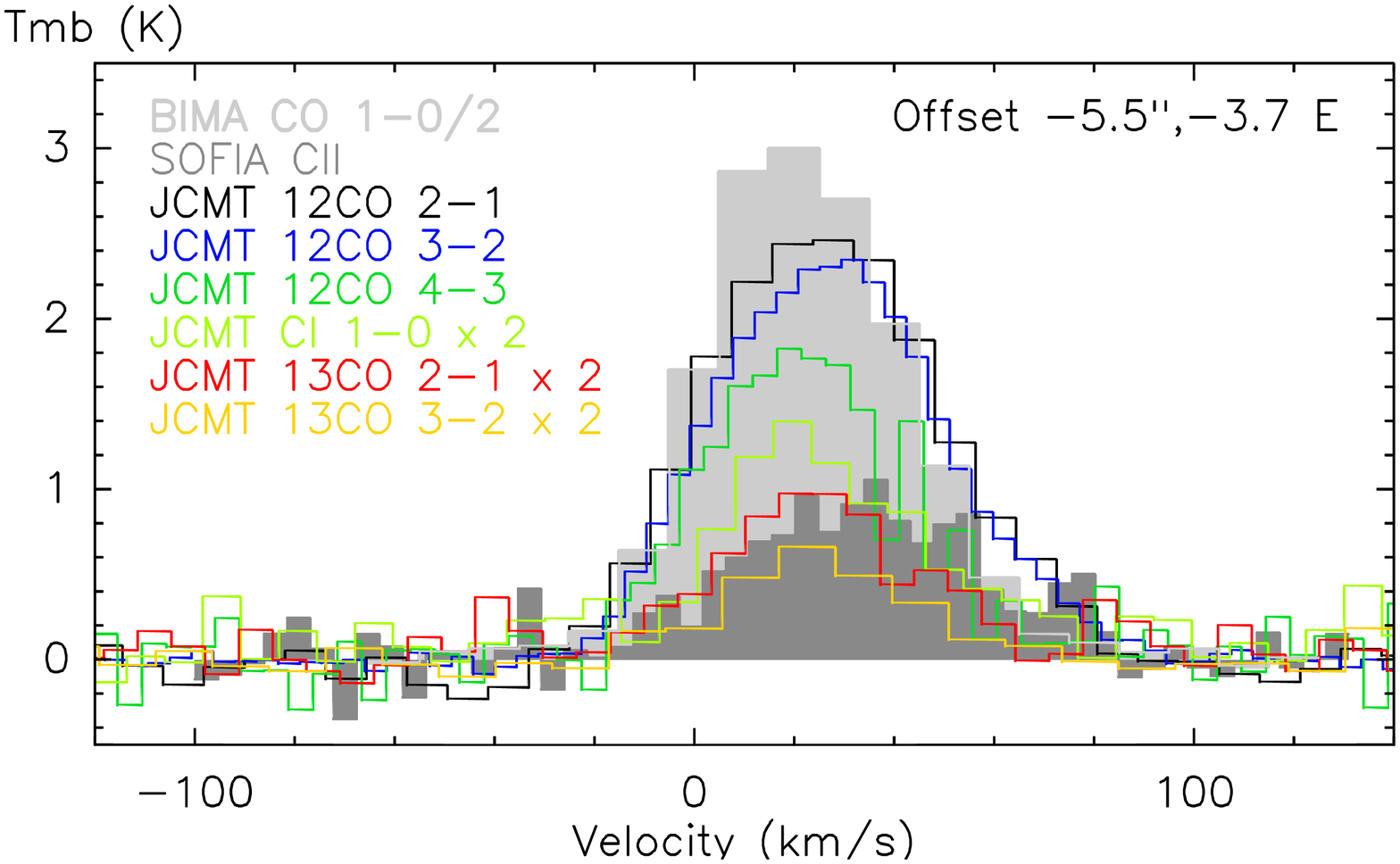,angle=0,width=0.85\linewidth} \\
 \end{tabular}

   \caption{\footnotesize{Averaged \cii\ spectra (dark gray, filled) toward the 
 GMC C (top), and GMC E (bottom) positions. The $^{12}$CO(2-1) spectra are divided by 2, the [CI]~$^3P_1-^3P_0$ and $^{13}$CO spectra are multiplied with 2. }}
   \label{fig:spectra}
 \end{figure}
In Fig.~\ref{fig:spectra} we show the \cii\  emission observed with GREAT (dark gray, filled) for both positions. On each spectrum we overlay  $^{12}$CO(1-0) data from BIMA-SONG\footnote{http://ned.ipac.caltech.edu/level5/March02/SONG/SONG.html}\citep{bimasong}, as well as $^{12}$CO(2-1),  $^{12}$CO(3-2),  $^{12}$CO(4-3),  $^{13}$CO(2-1),  $^{13}$CO(3-2), and [CI]~$^3P_1-^3P_0$ spectra from \citet{israel2003}. 
 Because of the different beam sizes, gridding, and map coverage it was not always possible to regrid and convolve all spectra to the positions and beam sizes of the GREAT observations.

%

Where possible, we smoothed the other data sets to the 15$''$ resolution of our \cii\ line observations.
We kept the $^{12}$CO(2-1) and $^{13}$CO(2-1) spectra on their native resolution of 22$''$ and 23$''$. We kept $^{12}$CO(3-2) and $^{13}$CO(3-2) on their native resolution of 15$''$.
 $^{12}$CO(1-0), $^{12}$CO(4-3), and  \ci $^3$P$_1$-$^3$P$_0$ spectra were smoothed to the resolution of our \cii\  data and interpolated at the two positions C and E.  
For the remaining ground-based spectra at the positions C and E  we averaged the data points closest to our \cii\ observations. 

The \cii\ emission is strongest at position E.
At both positions the different transitions show a good correlation of the central line velocities and line widths. 
The spectra at position C exhibit a consistent central velocity of ~40 \kms\ and similiar line widths of about 50 \kms. Only the \cii\ line has a broader line width and is blueshifted by a few \kms. 
Position E shows slight variations in the different spectral lines.  All lines have a comparable line width of slightly less than 50 \kms.   Most lines appear to be centered around 24 \kms, except for \twco(2-1) and (3-2) with slightly redshifted central velocities. 

We performed a Gaussian fit for all lines at the two positions to determine their respective line parameters such as peak- and line-integrated intensity  $T_{mb}^{peak}$ and $\int T_{mb}d\mbox{v}$, line width  $\Delta V$ , and line center velocity $\mbox{v}_0$. For position C we fitted a single-component Gaussian to the data without any additional constraints. The results are given in Table \ref{tab:1}. At position E we already noted the slight asymmetry in several of the CO lines, possibly resulting from two different, spatially and kinematically unrelated gas components. 
We used the optically thinnest lines from our data set, the \thco\ lines \citep{israel2003}, to specify the line center velocity of 24 \kms\ of the main component by fitting a single Gaussian to their profiles.
To quantify the contributions of two separate components to the spectra, we 
fitted 
two 
separate 
Gaussian components to the spectra while keeping the line center velocity of the main component fixed  to 24 \kms. Only for \twco(4-3) and \ci\ it was not possible to find a second line-component, therefore we performed an unconstrained single-component fit that we can attribute to the main component because of the comparable velocities. For \twco(1-0) we fixed the velocity and width of the high-velocity component to 53 and 47 \kms\ , respectively, and derived the line parameters for the low-velocity component. 
The decomposition into two different components is also supported by the BIMA \twco(1-0) data at 5.5$''$ resolution, which shows a line center velocity $\mbox{v}_0\approx50$~\kms\ at our center position. Therefore, emission from the PDR-dominated gas close to the central cluster is expected at $\mbox{v}_0\approx50$~\kms.

As an independant method to derive the \cii\ line parameters, we also calculated the zeroth, first and second moment of the spectra, i.e. the integrated intensity, mean velocity and line width.
The derived emission line parameters are summarized in Table \ref{tab:1} (moments are given as italic numbers).

The GMC E and B are both located southwest of the central position of IC~342 with an angular separation of only 3.5$''$. Accordingly, the 15$''$ \cii\ beam will always pick up emission of both GMCs. The \cii\ spectrum of position~E shows the strongest emission from both positions most likely because of the already known PDR/starburst contribution from GMC~B \citep{meier2005}. 
Position C is situated in the molecular ring, closest to the central cluster and will most likely contain contributions from dense molecular gas as well as from PDRs.  

\citet{israel2003} showed the existence of strong velocity gradients in the central region of IC~342. To quantify what pointing shift could alternatively mimic the observed line shift, we convolved the BIMA \twco\ (1-0) data to 15$''$. We found that a pointing offset of 7 $''$, significantly higher than the quoted accumulative pointing accuracy after 20 minutes, exactly in the direction of the central cluster could explain the \cii\ line profile. Therefore, it is unlikely, though it cannot be fully excluded, that a systematic pointing error may have caused the observed velocity shift.


   \begin{table}[!pt]
      \caption[]{IC~342 line parameters derived from Gaussian fits. One Gaussian component was fitted to all spectra at position C. At position E we fitted two Gaussian components to the spectra with a line center velocity of component 1 fixed to 24 \kms\ as derived from the single-component fit to the \thco(2-1) spectrum. 
The line parameters for \cii\ are additionally calculated from their moments between -30 and 100 \kms (numbers in italics).
}
         \label{tab:1}
         \centering
         \scriptsize

         \begin{tabular*}{0.48\textwidth}{@{\extracolsep{\fill}} l c c c c}

            \hline\hline
	    \noalign{\smallskip}
            Transition &$\int T_{mb}d\mbox{v}$& $\mbox{v}_0$&  $\Delta V$ & $T_{mb}^{peak}$\\
                       &[K \kms] & [\kms]&  [\kms]& [K]\\




             \noalign{\smallskip}
             \hline
	    
             \noalign{\smallskip}

             \multicolumn{5}{c} {Position C $(+3.6'', +3.6'')$} \\
             \noalign{\smallskip}
             \hline
             \noalign{\smallskip}

 		\cii\ &24.3$\pm$2.7 & 33.1$\pm$4.5 & 78$\pm$9 & 0.3 \\
                     \cii\ \tablefootmark{a} &\it{23.5} &\it{ 31.8} &\it{ 67.5} & \\ 
            
 	         $^{12}$CO(1-0)\tablefootmark{b} &307$\pm$2&41.5$\pm$0.2&50$\pm$1&5.83 \\
                  $^{12}$CO(2-1) &131$\pm$3&38.2$\pm$0.5&55$\pm$1&2.21 \\
 	         $^{12}$CO(3-2) &134$\pm$1&39.4$\pm$0.2&53$\pm$1&2.37 \\
 	         $^{12}$CO(4-3) &98$\pm$3&39.7$\pm$0.9&54$\pm$2&1.71 \\
 	         \ci $^3$P$_1$-$^3$P$_0$ &25.6$\pm$2.2&38.3$\pm$2.0&47$\pm$5&0.51 \\
           			
            \noalign{\smallskip}
            \hline
	    
            \noalign{\smallskip}

            \multicolumn{5}{c} {Position E $(-5.5'', -3.7'')$} \\
            \noalign{\smallskip}
            \hline
            \noalign{\smallskip}
		\cii\ &37$\pm$15&24\tablefootmark{c}&48$\pm$9&0.72 \\
                      &20.6$\pm$9.9&52.7$\pm$11.3&47$\pm$20&0.42 \\
            
                    \cii\ \tablefootmark{a} &\it{59} &\it{ 34.9} &\it{ 51} & \\
            
                 $^{12}$CO(1-0)\tablefootmark{b} &280$\pm$11&20&43$\pm$2&6.29 \\
                                &21.8$\pm$9.4&53\tablefootmark{d}&47\tablefootmark{e}&1.45 \\
                 $^{12}$CO(2-1) &139$\pm$6&24\tablefootmark{c}&50$\pm$2&2.59 \\
                                &9.5$\pm$4.2&59.8$\pm$4.5&34$\pm$11&0.26 \\
	         $^{12}$CO(3-2) &120$\pm$5&24\tablefootmark{c}&49$\pm$1&2.27 \\
                                &16$\pm$3&53.3$\pm$4.3&46$\pm$8&0.32 \\
	         $^{13}$CO(2-1) &23.6$\pm$2.7&24.0$\pm$1.7&48$\pm$4&0.46 \\
                                &-&-&-&- \\
	         $^{13}$CO(3-2) &13.9$\pm$2.0&24.4$\pm$1.8&41$\pm$4&0.32 \\
                                &-&-&-&- \\
	         $^{12}$CO(4-3) &88$\pm$5&21.2$\pm$0.9&46$\pm$2&1.81 \\
                                &-&-&-&- \\
	         \ci $^3$P$_1$-$^3$P$_0$  &34$\pm$4&24.4$\pm$1.7&51$\pm$4&0.61 \\            
					
            \noalign{\smallskip}
            \hline	    	    
	    
         \end{tabular*}
   \tablefoot{
              \tablefoottext{a}{Results from calculating the moments of the line.}
              \tablefoottext{b}{To convert from flux density Jy beam$^{-1}$km s$^{-1}$ to brightness temperatures we applied a factor of 4.4 assuming an effective beam size of 5.5$''$ \citep{bimasong}.}
              \tablefoottext{c}{The line center velocity was set to 24\kms.}
\tablefoottext{d}{The line center velocity was set to 53\kms. }
\tablefoottext{e}{The line width was set to 47\kms. }
              }

   \end{table}

\section{Discussion}
Table~\ref{tab:2}  lists line ratios of \cii\ and \ci\ to \twco(1-0) and \twco(2-1) and \twco(4-3)/\twco(1-0) (if available, the numbers for position E are given in the order high/low velocity component). Stronger PDR emission will be reflect in a higher ratio of the  fine structure lines to the molecular lines. Apparently, GMC~E shows the strongest PDR contribution. \citet{stacey1991} give a \cii/\twco(1-0) ratio of 5000 (corrected for main beam efficiency and assuming unity beam filling). They cite \cii\ intensities (beamsize 55$''$) of 3$\times$10$^{-4}$ erg~s$^{-1}$~cm$^{-2}$~sr$^{-1}$,
but only 40 \Kkms\ for the \twco(1-0) line intensity  in a 60$''$ beam. 
The \cii\ and \twco(1-0) data were taken using the Kuiper Airborn Observatory (KAO) and the Owens Valley Radio Observatory (OVRO), respectively. Given the larger beam sizes and unknown filling factors, their values are consistent with our data.

\citet{stacey1991} showed that  starburst galaxies have a high \cii/\twco(1-0) ratio of ~4100 while cooler, less active galaxies show much weaker ratios. Decomposition of the \twco(1-0) data into a low- and high-velocity component at 24 and 53 \kms\ , respectively, gives \cii/\twco(1-0) ratios of 600 and 4000, 
 indicating contributions from quiescent, weakly excited gas plus a strong, starburst like contribution from the local PDRs.

The \cii\ emitting level lies 91 K above the ground state and has a critical density of $\gtrsim 3500$~cm$^{-3}$.
Assuming optically thin emission and a level population in the high-temperature, high-density limit, we can calculate a lower limit of the C$^+$ column density from the integrated intensity \citep{crawford1985}. The results are 
given in Table~\ref{tab:2}. The beam-averaged column densities have  values between $0.9-1.6\times 10^{17}$~cm$^{-2}$, relatively low numbers for massive PDRs. 
The source intrinsic column density will be accordingly higher for a compact source, lower than the beam: a beam filling factor of 1/10, not unlikely given the small size of the cores visible in the interferometric \twco(1-0) map, will give a column density that is a factor of 10 higher.
Column densities of $1-2\times 10^{17}$~cm$^{-2}$ and a [C$^+$]/[H] ratio of 10$^{-4}$ gives a total mass in the 15$''$ beam of $5-10 \times 10^6$ M$_\odot$ at a distance of 3.9 Mpc.
\citet{israel2003} give comparable central masses of IC~342 of $5-7 \times 10^6$ M$_\odot$, given the uncertainties in the relative abundance of ionized carbon, filling factors, and distance.

\begin{table}
\caption{\label{tab:2} Line ratios and estimated column densities of the observed GMCs. For GMC E we give the values for the high/low velocity components. The PDR fit assumes a distance of 3.9 Mpc.}
\centering
\begin{tabular}{lcc}
\hline\hline
 \noalign{\smallskip}
& GMC C & GMC E\\
\hline
 \noalign{\smallskip}
\cii/\twco(1-0)&482&4236/692\\
\cii/\twco(2-1)&142&1223/150\\
\ci/\twco(1-0)&6.5&-/9.4\\
\ci/\twco(2-1)&1.9&-/2.4\\
\twco(4-3)/\twco(1-0)&20.7&21.1(-)\\
$N_\mathrm{[CII]}$/[10$^{17}$cm$^{-2}$]\tablefootmark{a}&1.4&0.9/1.6\\
            \noalign{\smallskip}
            \hline
            \noalign{\smallskip}

            \multicolumn{3}{c} {PDR model results} \\
            \noalign{\smallskip}
            \hline
            \noalign{\smallskip}
$\langle n \rangle$\,\,[10$^3$ cm$^{-3}$]&5.0&10/2.0\\
M$_\mathrm{tot}$\,\, [10$^6$ M$_\odot$]&20&2.0/15\\
$\chi$\,\, [Draine] &7&300/5\\
\hline
\end{tabular}
\tablefoot{
The line ratios were calculated from line-integrated intensities in units of erg s$^{-1}$ cm$^{-2}$ sr$^{-1}$.\\
 \tablefoottext{a}{Assuming a high-temperature and high-density limit and optical thin emission.}
} 
\end{table}
To confine the local gas parameters we used the KOSMA-$\tau$ PDR model code \citep{stoerzer1996, roellig2006} to model the emission of an ensemble of clumpy PDRs in a beam of 15$''$ \citep{cubick2008}. We fitted the model to the absolute intensities therefore making the total model gas mass sensitive to the distance of the PDR.
As a result from the fit we receive the mean gas density in the beam, the total gas mass, and the FUV intensity, that illuminates the PDRs in units of the Draine field \citep{draine1978}. The derived parameters for all positions are given in Tab.~\ref{tab:2} assuming a distance of 3.9 Mpc. Employing a shorter distance will reduce the total gas mass estimates accordingly.
 
To perform a consistent fit across both positions we excluded the \thco\ lines from the fitting. 
Note 
that the \thco\ lines show relative weak emission, which requires lower model densities of $\sim$1000~cm$^{-3}$ for both positions. Without fitting to the \thco\ lines the densities are higher by a factor of a few. Furthermore, the \cii\ and \ci\ line intensities relative to the CO emission can only be explained by densities below 10$^4$~cm$^{-3}$ and low FUV fields. Note that the model computations assume solar metallicities, while \citet{engelbracht2008} give somewhat higher values for IC~342. Higher metallicities will lead to lower \cii/CO ratios. 

The fit results 
from 
GMC E support the decomposition into two separate gas components. The lower velocity of 24 \kms\ would be consistent with assuming that the emission originates in the trailing, spiral arm region, which is connected to the molecular ring. The emission of the high-velocity component resembles the PDR signature of a starburst/PDR environment with mean FUV fields of 250-300 
(in units of the Draine field) 
and densities of 10$^4$ cm$^{-3}$. If we assume that the observations at position E were mispointed (see discussion in the previous section), we derive the following PDR model parameters:$\langle n\rangle=1000$~cm$^{-3}$, $\mathrm{M}_\mathrm{tot}=19\times 10^{6}$~M$_\odot$ and $\chi=5$. These gas parameters are in conflict with a scenario where a PDR/starburst dominates at the center of the galaxy.




\section{Conclusions}
We used the dual-band heterodyne receiver GREAT on board the airborne telescope SOFIA to observe two giant molecular clouds situated around the nucleus of IC~342 in the \twco~$J=11\to10$ transition and the \cii~158~\mum\ fine-structure line. We detected \cii\ emission at both positions but could not detect any \twco~$J=11\to10$ emission. 

The new SOFIA/GREAT spectra reveal a spectral distribution of the \cii\ emission that follows the distribution of the neutral and molecular gas, \ci\ and CO. 
The \cii\ spectrum observed at the position of GMC E shows 
two velocity components, a high-velocity component
that we attribute to emission from a PDR/starburst region in the molecular ring close to the central cluster with densities of 10$^4$ cm$^{-3}$, FUV field of 250-300 and a total mass of $2 \times 10^6$~M$_\odot$ 
and a
cooler, low-velocity component with densities of $2 \times 10^3$ cm$^{-3}$, FUV fields of a few and a total mass 7-8 times higher than the starburst component.
The model for GMC C gives  model parameters of densities of $5 \times 10^3$ cm$^{-3}$, FUV field of {\bf 7} and a total mass of $2 \times 10^7$~M$_\odot$.  

Despite the challenges that one might expect in an early transfer flight of SOFIA to Germany, we were able to deduce important astrophysical results primarily owing to the high spectral resolution available with the GREAT receiver.  These data demonstrate the promise of the GREAT/SOFIA facility for future work, such as detailed mapping of the \cii\ emission from the central regions of IC 342 and other nearby galaxies.


\acknowledgements{

We thank the SOFIA engineering and operations teams, whose tireless  
support and good-spirit teamwork has been essential for the GREAT  
accomplishments during Early Science, and say Herzlichen Dank to the
DSI telescope engineering team.

Based [in part] on observations made with the NASA/DLR Stratospheric Observatory for Infrared Astronomy. SOFIA Science Mission Operations are conducted jointly by the Universities Space Research Association, Inc., under NASA contract NAS2-97001, and the Deutsches SOFIA Institut under DLR contract 50 OK 0901.  

The research presented here was supported by the {\it Deutsche Forschungsgemeinschaft, DFG} through project number SFB956C.
}

\bibliographystyle{aa}
\setlength{\bibsep}{-2.1pt}
\bibliography{ic342}



\end{document}